\documentstyle[epsfig,graphicx,natbib,color,txfonts,longtable]{aa}
\begin{document}
\def\ovr{$\overline{R}$}
\def\ovv{$\overline{V}$}
\def\ovvi{$\overline{V}-\overline{I}$}
\def\ovbv{$\overline{B}-\overline{V}$}

    \title{The first search for variable stars in the open cluster NGC~6253 \\
           and its surrounding field
    \thanks{Based on observation made at the European Southern Observatory,
             La Silla, Chile, Proposal 073.C-0227.}$^,$
    \thanks{
Timeseries and light curves are available in electronic form
at the CDS via anonymous ftp to cdsarc.u-strasbg.fr (130.79.128.5).
or via http://cdsweb.u-strasbg.fr/cgi-bin/qcat?J/A+A/.
}}

   \author{
          F. De~Marchi
          \inst{1,2},
          E. Poretti
          \inst{3},
          M. Montalto
          \inst{4,5},
          S. Desidera
          \inst{6},
          \and
          G. Piotto
          \inst{1}
          }

   \authorrunning{De~Marchi et al.}
\titlerunning{Variable stars in NGC 6253}
   \offprints{F.~De~Marchi\\
              \email{fdemarchi@science.unitn.it} }

   \institute{Dipartimento di Astronomia, Universit\`a di Padova,
              Vicolo dell'Osservatorio 2, 35122 Padova, Italy
              \and
Dipartimento di Fisica, Universit\`a di Trento, Via Sommarive 14, 38123 Povo (TN), Italy
              \and
              INAF -- Osservatorio Astronomico di Brera,
              Via E. Bianchi 46, 23807 Merate (LC), Italy
              \and
              Universitaets-Sternwarte der Ludwig-Maximilians-Universitaet,
              Scheinerstr.~1, 81679 Muenchen, Germany
              \and
              Max-Planck-Institute for Extraterrestrial Physics,
              Giessenbachstr., Garching bei Muenchen, 85741, Germany
              \and
              INAF -- Osservatorio Astronomico di Padova,
              Vicolo dell'Osservatorio 5, 35122 Padova, Italy
             }

   \date{Received, accepted}

   \abstract
{}
{This work presents the first high--precision variability survey in the
 field of the intermediate-age, metal--rich open cluster NGC~6253. Clusters
of this type are benchmarks for stellar evolution models.}
{Continuous photometric monitoring of the cluster and its surrounding field
 was performed over a time span of ten  nights using the Wide Field Imager mounted
at the ESO-MPI 2.2m telescope. High--quality timeseries, each composed of about 800
datapoints, were obtained for 250,000 stars using ISIS and DAOPHOT packages.
Candidate members were selected  by using the 
colour--magnitude diagrams and period--luminosity--colour relations.
Membership probabilities based on the proper motions were also used.
The membership of all the variables discovered within a radius of 8\arcmin\, from the
centre is discussed by comparing the incidence of the classes in the cluster direction
and in the surrounding field.
}
{We discovered 595 variables and we also characterized most of them providing their 
variability classes, periods, and
amplitudes. The sample is complete for short periods: we
classified  20 pulsating variables, 225 contact systems, 99 eclipsing systems (22 $\beta$ Lyr type, 59
$\beta$ Per type, 18  RS CVn type), and 77 rotational variables.
The time--baseline hampered the precise characterization of 173 variables with periods longer than 4--5 days.
 Moreover, we found a cataclysmic system undergoing an outburst of about 2.5~mag.
We propose a list of 35 variable stars 
as probable members of NGC~6253.} 
{}


\maketitle

\section{Introduction}
\label{s:intro}

NGC~6253 and NGC~6791  are the only 
open clusters whose metallicities
above [Fe/H]=+0.3 were confirmed by spectroscopic analyses
\citep{carretta00,carretta07,sestito07}.
Therefore, these clusters are of special interest in several fields,
e.g., as  benchmarks for stellar evolution and stellar population models
and as targets for the search for extrasolar planets.
We observed both clusters in the framework  
of our project looking for transiting planets in super--metal--rich
open clusters. The results obtained on  NGC~6791 
were presented by \citet{montalto07}.

We also performed a 10-night observing campaign on NGC~6253 
with the same purposes as for NGC~6791.
In the first paper based on our new  investigation, \citet{montalto08}
obtained broad band photometry and astrometry for 187,963 stars
within 30 arcmin from the cluster.
Images from ESO archive \citep{momany01} were also used to
derive relative proper motions
and then distinguish between field stars and cluster members.
The availability of the astrometric cluster memberships and the photometric
quality of the new data allowed  new, independent determinations of the 
cluster's main parameters. 
Indeed, the determinations of the NGC~6253 parameters 
are affected by larger uncertainties because of  the cluster's projection toward a very rich
stellar field fairly close to the galactic centre
($l$=335.46~deg, $b$=--6.25~deg).  Systematic differences in the photometric 
calibrations of different datasets have been found 
\citep{bragaglia97, piatti98, sagar01, twarog03, anthonytwarog06}.
In this paper we adopt the values of the distance modulus
and of the reddenings  obtained by \citet{montalto08} using the technique
of the isochrone fitting,
i.e., $(m-M)_V$=11.68$\pm$0.10~mag, $E(B-V)$=0.15$\pm$0.02~mag and
$E(V-I)$=0.25$\pm$0.02~mag.
These values are also consistent with a weighted mean of all the
determinations.  The cluster age is about 3.5~Gyr
\citep{montalto08}.

Our project gives the possibility of studying
stellar variability in  super--metal--rich stars
using high--quality data \citep{demarchi07}. 
Since no variability survey on NGC~6253 has previously been performed,
we characterize   the  variable stars in NGC~6253 and in its
surrounding field for the first time. To do that, 
we started from the new findings and calibrations obtained by \citet{montalto08} 
so we refer the reader to that paper  
for a more detailed explanation of the methodologies applied
to determine the properties  and the fundamental parameters
of the cluster. 

\section{Observations and data reduction}
\label{s:obs}
\begin{figure}[]
\centering
\includegraphics[width=.99\columnwidth,height=0.99\columnwidth]{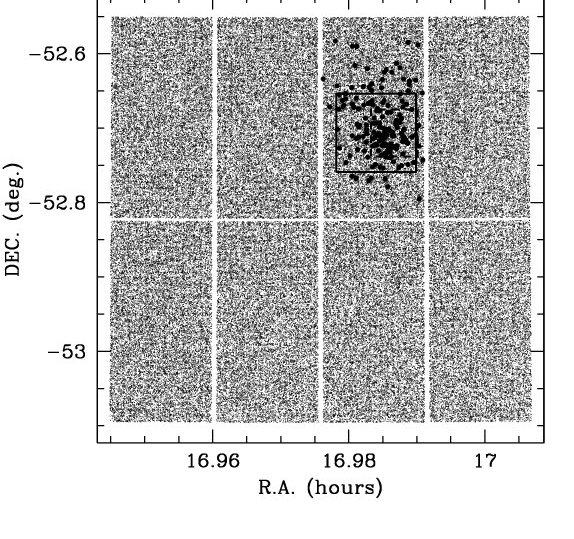}
\caption{\footnotesize Image of the WFI field (32~$\times$~32 arcmin$^2$).
 Solid lines represent the edges of the 6.3x6.3~arcmin$^2$ box surveyed  by 
 \citet{bragaglia97}. 
 Large points are stars with membership probabilities
(available only for stars located in chip~2) greater than 90\%.
 Chips are numbered from 1 (top right) to 8 (bottom right).}
\label{f:map_n6253}
\end{figure}
\begin{figure}[]
\centering
\includegraphics[width=0.95\columnwidth,height=0.95\columnwidth]{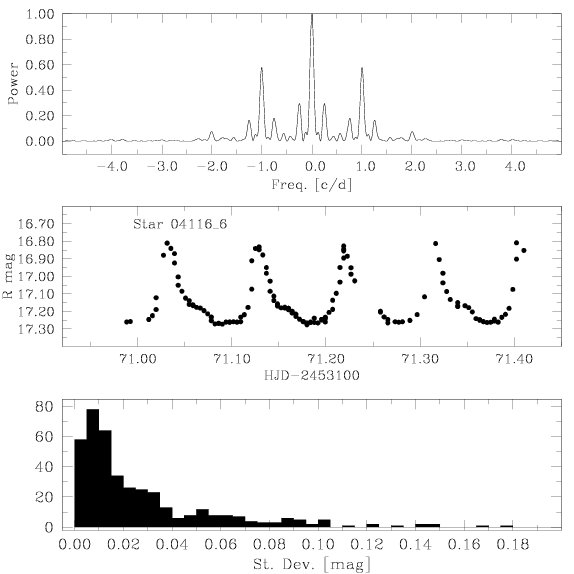}
\caption{ \footnotesize {\it Upper panel:} spectral window
of the timeseries of the variable stars in NGC~6253.
{\it Middle panel:} Example of an unfolded light curve: the high--amplitude
$\delta$ Sct star 044116\_6. {\it Bottom panel:} Histograms of the standard
deviations of the least--squares fit on the light curves of the periodic variables.
 }
\label{combi}
\end{figure}

NGC~6253 was observed for 10 consecutive nights
(from June~13, 2004 to June~22, 2004) using the
wide--field imager (WFI) mounted at the 
ESO-MPI 2.2m telescope, La~Silla, Chile.
 The WFI instrument includes  a mosaic
of eight  2k$\times$4k CCDs. The pixel scale is 0.238 arcsec/pixel.
In total, $\sim$45.3 hours of
observation were collected, mainly in the $R$ filter.
A few deep images in the $B$, $V$, and $I$ filters were also acquired to
construct colour--magnitude diagrams (CMDs), along with a
standard field to allow the calibration of the data.
In total 918 images of the cluster were obtained, with a mean exposure
time of 178 seconds. Table~\ref{t:obs} reports the journal of
observations and Fig.~\ref{f:map_n6253} shows a WFI image of NGC~6253.
Since the the size of each chip is  8\arcmin\, in right ascension   and 
16\arcmin\, in declination, we centered the cluster on one chip to minimize
the loss of stars between chips. 
Observations and data reduction to derive the calibrated photometry and the 
CMDs  of the cluster are described in more detail in \citet{montalto08}. The procedure 
to derive the light curves uses both ISIS~2.2 \citep{alard,alard2000} and DAOPHOT~II 
\citep{stetson} packages, as described in \citet{montalto07}. 

The length of the  observing nights (more than 0.32~d in 7~cases and more than 0.40~d in 5 cases,
 see Table~\ref{t:obs}) reduced the height of the aliases situated at $\pm$1~d$^{-1}$
from the central peak down to below 60\% of the power (Fig.~\ref{combi}, upper panel).
 Moreover, the light curves are very dense and
their shape clearly defined on each night (Fig.~\ref{combi}, middle panel).
Both these facts made the period detection quite straightforward, not only in
the case of high--amplitude variables, but most of time also for small--amplitude,
short--period variable stars.

As can be noted in Fig.~\ref{f:map_n6253}, our survey covers a much larger field of view
than the  previous ones (6.3x6.3~arcmin$^2$ by \citealt{bragaglia97}, 3.8x3.8~arcmin$^2$ by
\citealt{piatti98}). 
We could also identify new
variable stars in a wide part of the surrounding field.
The ISIS~2.2. and DAOPHOT~II packages returned
a photometric precision well below 0.01~mag
in the magnitude range $14\le R\le19$.  A plot of
the standard errors of the mean magnitudes in different filters
is shown in Fig.~1 in \citet{montalto08}.
Stars brighter than the turn-off magnitude ($V$=14.5)
are saturated in our photometry and cannot be
studied. In particular, this constraint hampers the
study of the variability of the
blue stragglers, as performed by \citet{thesis} in the more
favourable case of NGC~6791.
\begin{table}[]
\caption[]{The observation log for each night and limits of the
field of view.}
\centering
\begin{tabular}{ccc |  ccc}
\hline
\noalign{\smallskip}
Date & t$_{start}$      & $t_{end}$ & Date & t$_{start}$  & $t_{end}$  \\
\multicolumn{1}{c}{[Year 2004]} & \multicolumn{2}{c|}{[HJD-2453100]}&
\multicolumn{1}{c}{[Year 2004]} & \multicolumn{2}{c}{[HJD-2453100]}   \\
\noalign{\smallskip}
\hline
         \noalign{\smallskip}
June, 13-14  & 70.57 & 70.91 & June, 18-19  & 75.48 & 75.90 \\
June, 14-15  & 71.49 & 71.91 & June, 19-20  & 76.46 & 76.67 \\
June, 15-16  & 72.46 & 72.90 & June, 20-21  & 77.84 & 77.87 \\
June, 16-17  & 73.69 & 73.87 & June, 21-22  & 78.44 & 78.76 \\
June, 17-18  & 74.49 & 74.89 & June, 22-23  & 79.46 & 79.90 \\
         \noalign{\smallskip}
\hline
\noalign{\smallskip}
$\alpha_{min}$ & \multicolumn{2}{l}{16$^h$ 56$^m$ 41.6} &   
$\alpha_{max}$ & \multicolumn{2}{l}{17$^h$ 00$^m$ 24.7}     \\
$\delta_{min}$ & \multicolumn{2}{l}{-53$^\circ$ 05\arcmin 43.8\arcsec} &
$\delta_{max}$ & \multicolumn{2}{l}{-52$^\circ$ 33\arcmin 00.8\arcsec} \\
         \noalign{\smallskip}
         \hline
      \end{tabular}
     \label{t:obs}
\end{table}

\section{Cluster membership}
NGC~6253 is a relatively small cluster, but \citet{bragaglia97}
noticed the necessity of moving 8\arcmin\, from the
cluster centre to find a legitimate external field.
We followed this prescription and  we adopted the centre coordinates given by \citet{bragaglia97}.

The measured stars are indicated by small points, the $\sim$150 stars
with membership probability (hereafter MP) greater than 90\% are highlighted with larger black points.
MPs were calculated in \citet{montalto08} following the approach proposed by \citet{mprob}:
\begin{equation}
MP=\Phi_c/(\Phi_c+\Phi_f)
\end{equation}
where $\Phi_c$ and $\Phi_f$ are the distribution of cluster and field stars
in the diagram of the proper motions, respectively. These distributions are
typically represented as Gaussian functions. The distribution of the
cluster stars has a narrow peak centered at $\mu_\alpha=\mu_\delta=0$, while the distribution
of field stars is much broader. For the given candidate member, the 
calculation of the MP was performed by selecting a surrounding sample of
a 2.5~mag range centered on the candidate's position. In such a way the local sample
stars compensate for the effect of a magnitude dependence of the cluster--to--field
star ratio. When constructing a $V-MP$ diagram, the stars belonging to the cluster
occupy a well--defined region (see Fig.~4 in \citealt{montalto08}). We require
the probable member clusters to have MP$>90\%$ at $V=12.5$ and MP$>50\%$ at $V=18.0$.

Since the determination of the MPs is a differential process and  the cluster is almost completely 
included in chip~2, the MPs are reliable only for stars 
belonging to this chip and brighter than $V$=18. Looking at the distribution of the stars with a high MP we can
infer that some members of the cluster might also be present in chips~1 and 3.

\section{The variable stars}
\label{s:var}
\subsection{Detection}
The ISIS~2.2 and DAOPHOT~II packages allowed us
to extract the first list of suspected variable stars from
the full database of  250,000 timeseries.
This list was
validated and shortened by calculating the parameters related to the reduction of the
initial variance obtained by introducing trial periodic terms.  
These parameters are the reduction factor \citep{vani} and the coefficient
of spectral correlation \citep{mello}. 
As in the case of NGC~6971 \citep{demarchi07},
we could separate short- and long- period variable stars by introducing a parameter
that is more sensitive to the night--to--night variations. 
Tests on the significance of the detected periodicities (e.g., signal--to--noise
ratio above 4.0 in amplitude) allowed us to get a more defined sample
of real variable stars. 
A few objects whose variability appears to stem from photometric
artefacts (e.g. eclipse-like features occurring exactly at the same time on the
second night) were removed from the list. These spurious photometric
effects are usually corrected when applying to the light-curve
algorithms such as the one developed by \citet{tamuz05}.
However, we noticed that the application of this algorithm 
degrades  the precision of the variable star photometry.
Therefore, being interested in much greater light  variations
than the tiny photometric effect of a planetary transit,
we decided to analyse the light curves before
applying  the algorithm.

We identified 595 variable stars at the end of our process, whose 
timeseries are composed of about 800~datapoints. 
To identify them we used the five--digit number assigned by our customized package 
package, followed by the number of the chip that the star belongs to. 
The timeseries are available at the ``Centre de Donn\'ees astronomiques de Strasbourg" (CDS).

\begin{table}[]
   \caption[]{Inventory  of the variables found in NGC~6253 and its surrounding area.
 }
\centering
       \begin{tabular}{lcccc}
         \hline
         \noalign{\smallskip}
     Type   &\multicolumn{4}{c}{Number of variables}\\
            & all chips   & $r<$8\arcmin & Candidate & Probable \\
            &             &              & members   & members \\
         \noalign{\smallskip}
         \hline
         \noalign{\smallskip}
RR~Lyrae       & 4   & 1          &  0   & 0 \\ 
$\delta$ Scuti & 11  & 0          &  0   & 0   \\ 
$\beta$ Cep    & 1   & 0          &  0   & 0   \\ 
HADS           & 4   & 0          &  0   & 0   \\ 
EW-type        & 225 & 50         &  16  & 8   \\ 
EB-type        & 22  & 6          &  2   & 0   \\ 
EA-type        & 59  & 13         &  5   & 1   \\ 
RS~CVn         & 18  & 4          &  2   & 1   \\ 
U~Geminorum    & 1   & 1          &  1   & 1   \\ 
Rotationals    & 77  & 27         &  16  & 15   \\ 
Long period    & 173 & 41         &  16  &  9   \\ 
         \noalign{\smallskip}
         \hline
      \end{tabular}
     \label{t:class_n6253}
\end{table}
\subsection{Classification}
The timeseries of the 595 variable stars were analysed in frequency
by using the least--squares iterative sine--wave search
\citep{vani} and the Phase Dispersion Minimization \citep{pdm} methods. 
The periods were refined by means of a least--squares procedure (MTRAP, \citealt{mtrap});
their error bars are in the range $1-6~\times~10^{-5}$~d. The bottom panel of
Fig.~\ref{combi} shows the distribution of the standard deviations
of the least--squares fits, indicating a median precision of 0.015~mag. 

We could show  amplitudes of light variability 
down to the 0.01~mag level. At this level, 
rotational variables could be separated from pulsating variables on the basis
of the period values and of the Fourier parameters alone \citep{ogle}.
On the other hand, it is very difficult to disentangle rotational from 
eclipsing variables. 
To distinguish  
rotational variables from  contact binaries, we referred to the degree
of asymmetry of the double--wave light curves and to the occurrence of the
minima at phases 0.00 and 0.50.  Of course,  we cannot rule out
that a small fraction of the variables classified as rotational variables might
be actually contact systems showing grazing eclipses or viceversa.

We considered
two classes of rotational  variables, RO1 and RO2 stars. RO1 stars show a light curve 
characterized by a single wave, which is often asymmetrical. RO2 stars
show a more complicated curve composed of two waves having unequal amplitude
and duration. This light curve is comes from two (groups of) spots located at different
latitudes that remain visible to the observer during  different fractions of the
rotational period. In some cases these spotted stars are observed in eclipsing 
systems, the so-called RS CVn variables. Other cases of eclipsing systems
are contact (W~UMa variables, EW), semi-detached ($\beta$ Lyr variables, EB), and
detached systems (Algol variables, EA) binaries.  In some cases it was very
difficult to distinguish between EW system showing grazing eclipses 
and rotational variables. 
We also identified three different classes of 
pulsating variables, i.e., RR Lyr, $\delta$ Sct and High--Amplitude $\delta$ Sct (HADS) 
stars. In both cases, eclipsing binaries and pulsating variables,  
the very good spectral window (Fig.~\ref{combi})  made the 
period detection quite straightforward. 
On the other hand, defining the periods longer than 4--5~d was not easy.
In particular it was impossible for periods longer than 10~d and we simply classify
these stars as long period (LON) variables. These stars are mostly rotational 
variables.
The  summary classification of the entire sample is reported
in Table~\ref{t:class_n6253}. Tables~\ref{tab_app_pul_1},
\ref{tab_app_EW}, \ref{tab_app_EA}, \ref{tab_app_EB}, \ref{tab_app_RSCVn},
\ref{tab_app_RO1}, \ref{tab_app_RO2}, and \ref{tab_app_LON}
list the members of each class giving  the identifier in the \citet{montalto07} catalogue,
the coordinates, the photometry, the epoch of maximum or minimum brightness (HJD--2453100), the period, the amplitude, 
the distance, and the MP values.
Uncertain MP values (stars with $V>$18, often close to the chip borders) are marked
with an asterisk.
The catalogue of the light curves of the periodic variables is available at the CDS.

We paid particular attention to the variables located
within 8\arcmin\, from the cluster centre;
in any cases no variable with MP larger than 50\%
 was found  at a greater distance.
If the MP is not available (stars near the edges of  chip~2
or stars fainter than $V$=18), the membership is estimated
from their location on the $B-V$ vs. $V$ and $V-I$ vs. $V$ CMDs.
Moreover, for pulsating variables and  contact binaries with
unknown membership, our conclusions are based on the applications
of the usual period--luminosity ($P-L$) and  period--luminosity--colour
($P-L-C$) relations.

\begin{figure*}[]
\centering
\includegraphics[width=0.8\columnwidth,height=0.8\columnwidth]{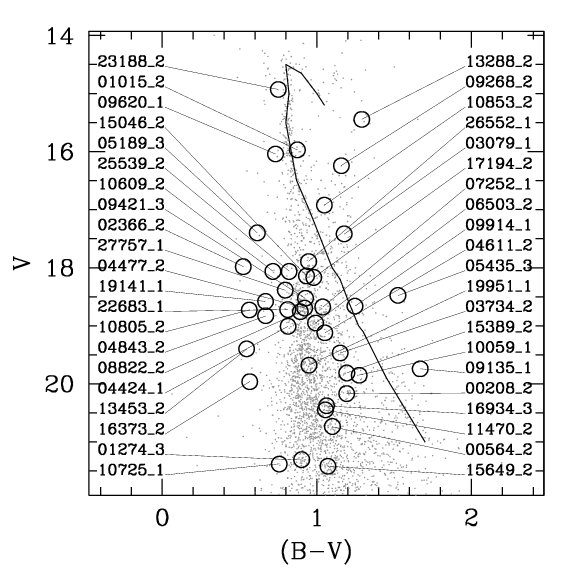}
\includegraphics[width=0.8\columnwidth,height=0.8\columnwidth]{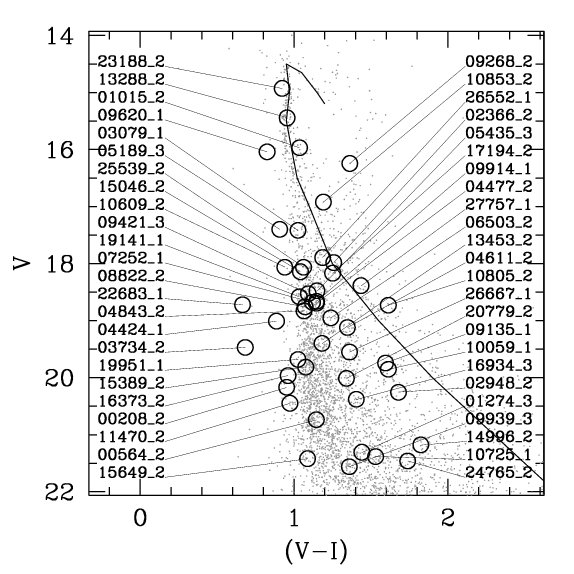}
\includegraphics[width=0.8\columnwidth,height=0.8\columnwidth]{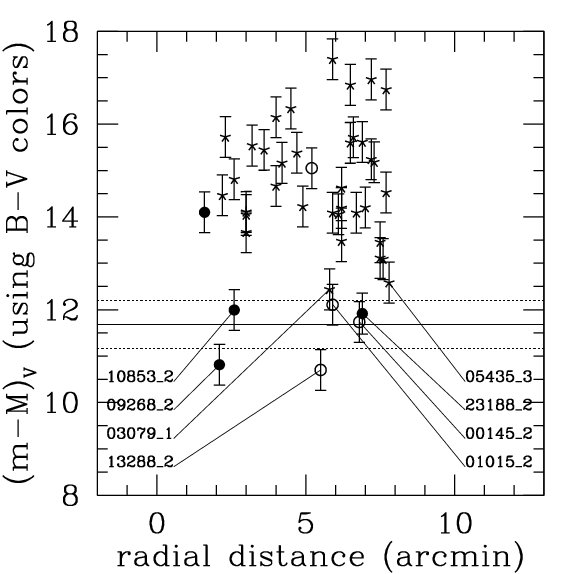}
\includegraphics[width=0.8\columnwidth,height=0.8\columnwidth]{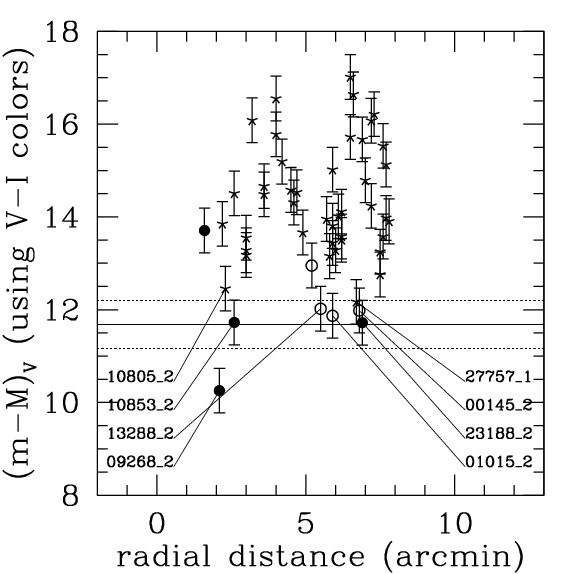}
\caption{ \footnotesize {\it Top row:}
   colour--magnitude diagrams of NGC~6253 with the contact binaries at $r<$8~\arcmin\, highlighted.
The Main Sequences are individuated by fiducial lines.
{\it Bottom row:} 
Distance moduli of all contact binaries at $r<$8\arcmin\,
  obtained using the $P$-$L$-$C$ relations.
We use both $(B-V)$ (left panel) and $(V-I)$ colours (right panel).
The horizontal line represents the distance modulus of the cluster resulting
from isochrone fitting \citep{montalto08}. Filled circles show the binaries with
MPs$>$50\%; open circles the stars with MPs$<$50\%, starred points the
binaries with unknown membership. The error bars are the errors associated
with the $M_V$ calculation and include errors in the colour determinations.}
\label{fig_hr_bv_EW_n6253}
\end{figure*}

\begin{figure*}[]
\centering
\includegraphics[width=0.6\columnwidth,height=0.6\columnwidth]{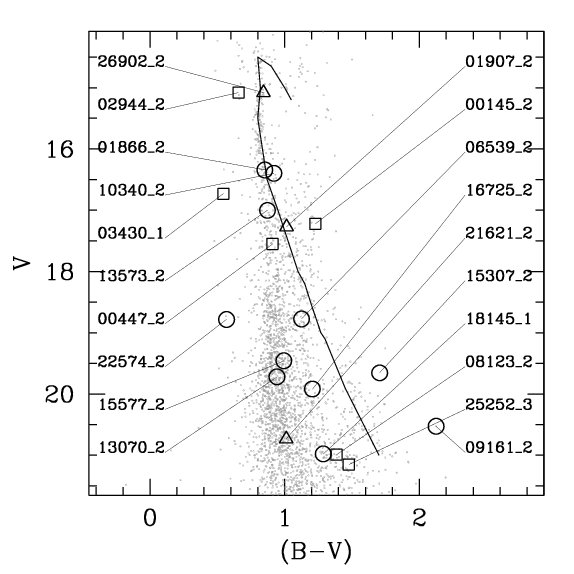}
\includegraphics[width=0.6\columnwidth,height=0.6\columnwidth]{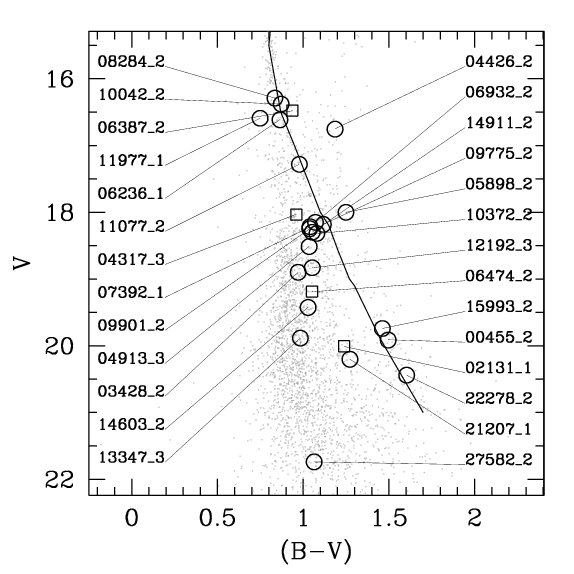}
\includegraphics[width=0.6\columnwidth,height=0.6\columnwidth]{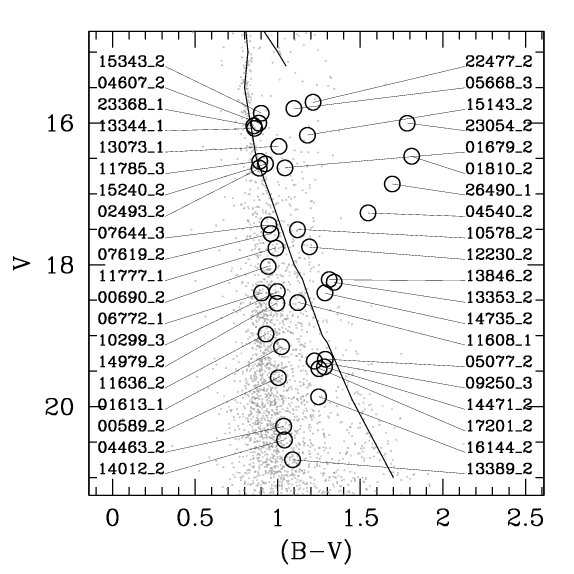}
\includegraphics[width=0.6\columnwidth,height=0.6\columnwidth]{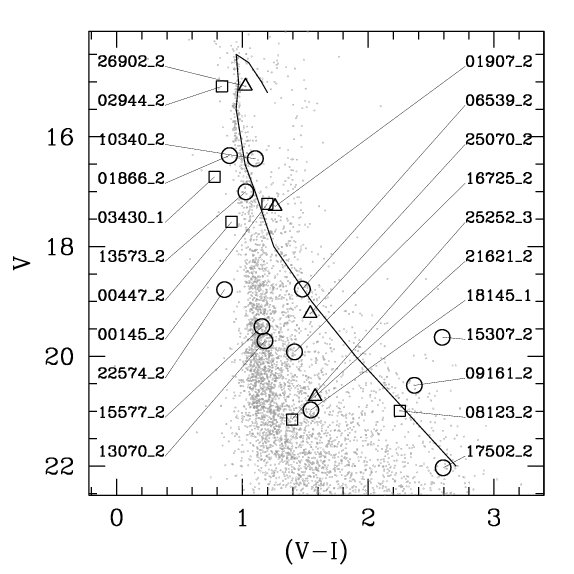}
\includegraphics[width=0.6\columnwidth,height=0.6\columnwidth]{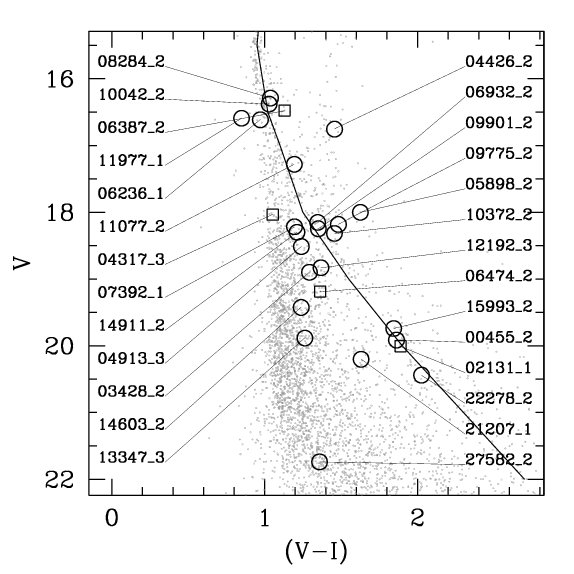}
\includegraphics[width=0.6\columnwidth,height=0.6\columnwidth]{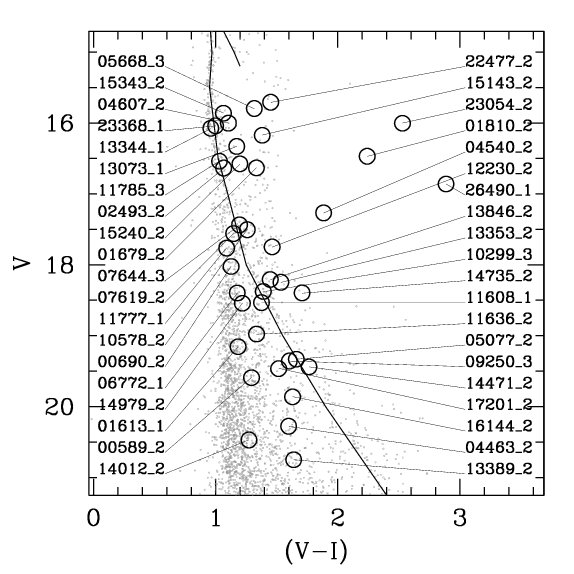}
\caption{ \footnotesize {\it Left panels:} CMDs of NGC~6253 with the EA (circles),
  EB (squares), RS~CVn (triangles) variables highlighted.
{\it Middle panels:}  CMDs of NGC~6253 with the rotational single-wave
 (circles), and double-wave (squares) variables highlighted.
{\it Right panels:} CMDs of NGC~6253 with the
long-period variables located within the 8\arcmin-radius circle highlighted.
The Main Sequences are individuated by fiducial lines.}
\label{fig_hr_ecl_n6253}
\end{figure*}

\subsection{Pulsating variables}

The only pulsating variable located at less than 8\arcmin\, from the centre
is the RRab star 10540\_2.
Its MP is quite high, but it is clearly too faint ($V$=17.39) to belong to the cluster.  
Amongst the other four RR Lyr stars, 
15578\_7  is a new galactic Blazhko variable.

Twelve variables show an amplitude smaller than 0.06~mag; since they have
a very short period (less than 0.10~d), we can rule out their being rotational variables.
All the $B-V$ values except one range from 0.42 to 0.79 mag, mostly between
0.50 and 0.63. This interval, taking the reddening $E(B-V)$=0.15~mag into account,
suggests their classification as $\delta$ Sct stars.  
The remaining low-amplitude variable shows $(B-V)=0.084$: it  probably
belongs to the $\beta$ Cep class. 

Four variables show a larger amplitude (more than 0.09~mag) and the asymmetric shape of
the light curve  typical for high--amplitude $\delta$ Sct stars. By using the
new period--luminosity relation derived by \citet{fornax}, 
no doubt is left on the fact that all these variables do not belong to NCG~6253.
Finally, none of the pulsating variables is a  member of NGC~6253, since they are
all located well beyond the cluster.

\subsection{Contact binaries}

We have 50 contact binaries located within the 8\arcmin\, radius and
the  $(B-V)$ and $(V-I)$ colours are both available for 44 binaries,
while only the $(V-I)$ colours are available for 6 of them.
For these stars it is possible to apply the $P-L-C$ relations
given by \citet{rucinski03}
and compare the resulting distance moduli with that of the cluster,
obtained by isochrone fitting \citep{montalto08}.
The errors on the distance moduli are calculated by considering the
uncertainties on the mean $B-V$ and $V-I$ colours.
It must also be  taken into account that, since our light curves are in the $R$ band,
 we cannot know the exact value $V_{max}$ of the magnitude at maximum
 brightness required by the \citet{rucinski03} calibrations.
We estimated  $V_{max}$ as $(V_{mean}-R_{mean})$+$R_{max}$; i.e., we assumed
that the colour of these binary systems does not change during the orbital period
since the components have a very similar temperature.

An estimate of the membership of the objects can be obtained using the CMDs (Fig.~\ref{fig_hr_bv_EW_n6253},
top row), the  MPs based on the proper motions, and the two $P-L-C$ relations
(Fig.~\ref{fig_hr_bv_EW_n6253}, bottom row). 
The fiducial lines shown in the CMDs are obtained by selecting 10--15 points at different
magnitudes along the Main Sequences, both from the observations described here and
from \citet{bragaglia97}.
We could select a list of 13 candidate members  
for which one of the above criteria is satisfied (Table~\ref{tab_EW_n6253}).

We note that 23188\_2 and 10853\_2 satisfy all the membership criteria
and then are very likely cluster members. 
The position in the CMDs and the $P-L-C$ relations also suggest
that 01015\_2 is a cluster member, but this hint is not supported by the MP,
which is very small.
The case of 09268\_2 is the opposite: it has also a fairly large MP, 
but the other indicators
suggest that is is more probably located between the Sun and the cluster.
Unfortunately, none of the remaining cases gives us enough confidence on a
cluster membership. 

We can tackle the problem of cluster membership in an indirect way.
In the surrounding field we found 175 EW binaries,
with an incidence of 0.21~EW\,arcmin$^{-2}$. Therefore, we should have 42 field
EW--stars superposed on the cluster. Since we found 50 stars (Table~\ref{t:class_n6253}),
the excess is only marginally significant.  
We have only two well--established memberships; therefore,  
we can reasonably estimate that very few contact binaries (up to six) among the 
remaining 11 candidates listed in Table~\ref{tab_EW_n6253}
actually belong to NGC~6253. This clue is confirmed by the candidates
do not match the photometric criteria very well (Table~\ref{tab_EW_n6253}).
In NGC~6791 we found three well-established and five likely EW--members  
 \citep{demarchi07}, i.e., similar countings. The surveys of the two clusters
are complete both at the magnitude and at the periods of the EW binaries. 
The two clusters have a different stellar content, since 
NGC~6253 has about 500-1000 members \citep{montalto08},
and NGC~6791  about 4900$\pm$1000 \citep{thesis}.
The similarity between EW countings in the two clusters
supports the hypothesis of an anticorrelation between the
frequency of binaries  and the richness of the host cluster \citep{K95}. 

Among the non-member contact binaries, we note
that 00441\_4 has a period of 0.21002~d, shorter than
the shortest contact binary found 
in the ASAS database \citep[$P$=0.217811~d, ASAS 083128+1953.1,]
[] {rucinski07}
and very similar to the binary with the shortest period known
\citep[$P$=0.2009~d, V344 in the Lupus field,][]  
{weldrake08}.

\subsection{Semi-detached and detached systems}
The sample of the semi-detached and detached systems
within 8\arcmin\, is composed of five EA, two EB, and
two RS~CVn stars (lower part of Table~\ref{tab_EW_n6253}). 
Their periods are shorter than 2.3~d.
The star 26902\_2 has a high MP, and it is the only case for which we can be
very confident about its membership, also confirmed by the positions
in the CMDs (Fig.~\ref{fig_hr_ecl_n6253}, left panels). On the basis
of the same criteria, 10340\_2 is  another probable member. On the
other hand, the MP value rules out the membership of 00145\_2. No
firm conclusion on the membership  can be drawn on the other cases.

\subsection{Rotational and long-period variables}

A great number of the new variables discovered in
our survey shows the  single (RO1) or double (RO2)  wave light
curves  typical of  rotational effect.  The 10--d time baseline
allowed us to detect all the variables with rotational periods
shorter than 4--5 days. Other variables 
show an evident  night--to--night variability, but we cannot
infer any reliable value for the period. These variables are
probably long-period ones (LON). 

By adopting the same criteria as used in other cases, we selected
the RO1 (14 stars),  RO2 (2 stars), and LON (16 stars)
candidate members of NGC~6253 (Table~\ref{tab_rot_n6253}). 
Figure~\ref{fig_hr_ecl_n6253} plots the CMDs with
the positions of all the rotational variables
within 8\arcmin~ from the centre and the positions of the long-period variables
highlighted (middle and right panels, respectively).

We discovered 27 variables
in the 8\arcmin~ radius from the centre and 16 of them can be considered 
candidate members on the basis of the positions on the CMDs and of the MPs
(upper part of Table~\ref{tab_rot_n6253}). 
The stars  10042\_2,  11077\_2, and 06387\_2 have a large 
MP and also a position on the CMDs compatible with cluster membership.
We count 57 rotational variables in the surrounding field, 
i.e., an occurrence
of 0.06~star\,arcmin$^{-2}$. This would imply an estimate of 12 field rotational 
variables along the line of sight of NGC~6253. There is a significant
difference between the expected and the observed number of rotational variables, and we
can infer that several selected candidate members (up to 15)
actually belong to the cluster. Considering the short periods of these 
stars and the old age of the cluster, it is likely that
the rotational variables that are cluster members are close, tidally locked
binaries.

In the same manner, we can estimate 32 LON field variables superposed to NGC~6253.
In turn this  means that  up to 9 out of the 41 LON variables discovered 
in the 8\arcmin\, radius can be considered  members of the cluster. These 9 stars 
should be found among the 16~candidate members listed in  Table~\ref{tab_rot_n6253}.

\begin{figure}[]
\centering
\includegraphics[width=.9\columnwidth,height=0.6\columnwidth]{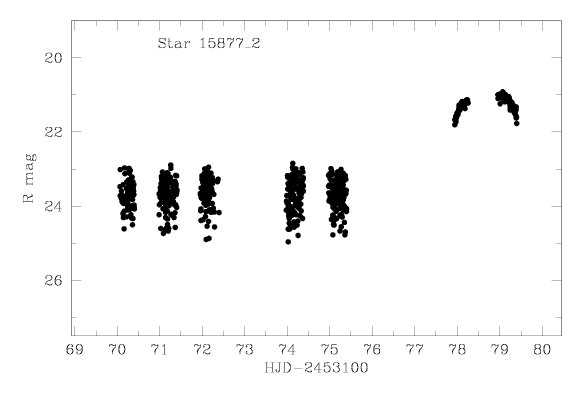}
\caption{ \footnotesize   Light curve of the new U~Gem cataclysmic variable.}
\label{f:cata_n6253}
\end{figure}
\subsection{A new  cataclysmic variable}
The U~Geminorum variable 15877\_2 is located at 6.8\arcmin\, from the cluster 
centre, but unfortunately its  MP is not available (Table~\ref{tab_rot_n6253}). 
In the light curve,
the scatter at the quiescence phase suggests some photospheric activity, but
no periodicity is detected by analysing these measurements.
Such a phase lastes the first 8 days of our survey. After that,
its brightness in the $R$-band increases by about 2.5~mag
(Fig.~\ref{f:cata_n6253}). 
The maximum is not observed because it occurred in daytime.
The star 15877\_2  
appears to be  similar to the U~Geminorum variable 06289\_9, classified
as a  member of NGC~6791 \citep{demarchi07}.

\section{Conclusions}
\label{s:discussion}

In this paper we have described the first search for 
variable stars in the open cluster NGC~6253. 
Since the membership probabilities based on the proper motions
are not reliable for stars with $V>18$, only a few variables 
could be confirmed directly as cluster members.
However, the 
comparison with the number of contact binaries and rotational 
variables (both short and long periods) found in a large area surrounding  
the cluster allowed us to estimate the incidence of these variables within the
cluster, too. On the basis of these considerations we propose 35 members
of NGC~6253 within the sample of variable stars,
though new observations are needed to identify
some of them in an unambiguous way.

The class of main--sequence rotational variables is the most numerous, as 
observed in the surrounding field. On the basis of  similar observing campaigns,
we found the same number of contact binaries in NGC~6253 as were previously found in NGC~6791, 
thus confirming the anticorrelation between the frequency of binaries and the richness
of the cluster \citep{K95}. This anticorrelation is similar to the one  
found between  the frequency of blue stragglers and
the total magnitude of the host cluster.
Both these facts can lead back to the important effects caused by mass loss 
in the evolution and in the  history of the dynamics of open clusters \citep{davies,fdm}.

We discovered a new eruptive variable in NGC~6253. A single outburst was observed, so
we cannot infer any physical characteristic of the system. Since we made the same discovery
in NGC~6971 \citep{demarchi07}, it seems that continuous surveys on a few nights are
very effective in finding these rare and interesting objects.

\begin{acknowledgements}

This work was funded by COFIN 2004
``From stars to planets: accretion, disk evolution and
planet formation'' by MIUR and by  PRIN 2006
``From disk to planetary systems: understanding the origin
and demographics of solar and extrasolar planetary systems''
by INAF. We thank the anonymous referee for careful reading
and useful suggestions, and J.~Vialle for checking the English form.

\end{acknowledgements}

\bibliographystyle{aa}

\clearpage

\begin{appendix}
\label{appA}
\section{
 Tables}
This appendix includes the Tables listing all the variables
discovered in our survey of NGC 6253 and its surrounding field.
The epochs of maximum or minimum brightness are expressed as HJD$-$2453100
in the columns $T_{\rm max}$ and $T_{\rm min}$.
\begin{enumerate}
\item Pulsating variables: Table~\ref{tab_app_pul_1};
\item EW--type variables: Table~\ref{tab_app_EW};
\item EA--type variables: Table~\ref{tab_app_EA};
\item EB--type variables: Table~\ref{tab_app_EB};
\item RS--CVn variables: Table~\ref{tab_app_RSCVn};
\item Rotational single--wave variables: Table~\ref{tab_app_RO1};
\item Rotational double--wave variables: Table~\ref{tab_app_RO2};
\item Long--period variables: Table~\ref{tab_app_LON}.
\end{enumerate}

The binary systems considered as candidate members of NGC~6253 are listed in
Table~\ref{tab_EW_n6253}.
The rotational and long-period variables
considered as candidate members of NGC~6253 are listed in 
Table~\ref{tab_rot_n6253}.

\begin{table*}
\begin{flushleft}
 \caption{\footnotesize Pulsating variables.}
\resizebox{0.99\textwidth}{!}{
\begin{tabular}{lrl ll cccc clcr r }
\hline \hline
\noalign{\smallskip}
\multicolumn{1}{c}{Star}& \multicolumn{1}{c}{ID} &
\multicolumn{1}{c}{Type}&
\multicolumn{1}{c}{$\alpha_{2000}$}&
\multicolumn{1}{c}{$\delta_{2000}$}&
\multicolumn{1}{c}{\ovr}&
\multicolumn{1}{c}{\ovv}&
\multicolumn{1}{c}{\ovbv}&
\multicolumn{1}{c}{\ovvi}&
\multicolumn{1}{c}{$T_{\rm max}$}&
\multicolumn{1}{c}{Period}&
\multicolumn{1}{c}{Ampl.}&
\multicolumn{1}{c}{Distance}&
\multicolumn{1}{l}{MP} \\
& & &\multicolumn{1}{c}{[deg]} &\multicolumn{1}{c}{[deg]} &\multicolumn{1}{c}{[mag]} &\multicolumn{1}{c}{[mag]} &\multicolumn{1}{c}{[mag]} &\multicolumn{1}{c}{[mag]}
&\multicolumn{1}{c}{[HJD]} &\multicolumn{1}{c}{[d]} &\multicolumn{1}{c}{[$R$ mag]} &\multicolumn{1}{c}{[arcmin]} &\multicolumn{1}{c}{\%}\\
\hline
\noalign{\smallskip}
\noalign{\smallskip}
16334\_6     &        132171     &     DSCT        &         254.517933029     &         -52.919478492     &     17.548      &     17.854      &     0.539       &     0.767       &     70.605      &     0.03065     &     0.02        &     15.6        &              -     \\   
11353\_4     &         79692     &     DSCT        &         254.373216875     &         -52.592925272     &     18.915      &     19.257      &     0.557       &     0.703       &     70.586      &     0.03818     &     0.04        &     16.0        &              -     \\   
04756\_6     &        123786     &     DSCT        &         254.547272025     &         -53.053321129     &     16.815      &     17.179      &     0.568       &     0.762       &     70.586      &     0.04065     &     0.04        &     22.2        &              -     \\   
22953\_1     &         16369     &     DSCT        &         255.057077875     &         -52.629821132     &     15.968      &     16.179      &     0.517       &     0.566       &     70.613      &     0.04182     &     0.02        &     11.4        &              -     \\
\hline \hline
\end{tabular}
}
\label{tab_app_pul_1}
\end{flushleft}
\end{table*}

\begin{table*}
\begin{flushleft}
 \caption{\footnotesize  Contact binaries, W Ursae Maioris (EW) systems.}
\resizebox{0.99\textwidth}{!}{
\begin{tabular}{lr ll cccc clcr r}
\hline
\hline
\noalign{\smallskip}
\multicolumn{1}{c}{Star}& \multicolumn{1}{c}{ID} &
\multicolumn{1}{c}{$\alpha_{2000}$}&
\multicolumn{1}{c}{$\delta_{2000}$}&
\multicolumn{1}{c}{$R_{\rm max}$}&
\multicolumn{1}{c}{\ovv}&
\multicolumn{1}{c}{\ovbv}&
\multicolumn{1}{c}{\ovvi}&
\multicolumn{1}{c}{$T_{\rm min}$}&
\multicolumn{1}{c}{Period}&
\multicolumn{1}{c}{Ampl.}&
\multicolumn{1}{c}{Distance}&
\multicolumn{1}{l}{MP}\\
& &\multicolumn{1}{c}{[deg]} &\multicolumn{1}{c}{[deg]} &\multicolumn{1}{c}{[mag]} &\multicolumn{1}{c}{[mag]} &\multicolumn{1}{c}{[mag]} &\multicolumn{1}{c}{[mag]}
&\multicolumn{1}{c}{[HJD]} &\multicolumn{1}{c}{[d]} &\multicolumn{1}{c}{[$R$ mag]} &\multicolumn{1}{c}{[arcmin]} &\multicolumn{1}{c}{\%}\\
\hline
\noalign{\smallskip}
00441\_4     &         71609     &         254.249749852     &         -52.783032714     &     21.373      &     22.350      &     --          &     --          &     70.586      &     0.21002     &     0.38        &     19.5        &              -     \\   
12192\_1     &          8870     &         255.062764181     &         -52.626856921     &     20.939      &     21.825      &     1.257       &     1.514       &     70.664      &     0.22407     &     0.33        &     11.7        &              -     \\   
10086\_5     &        103181     &         254.190530241     &         -52.847147618     &     21.454      &     22.485      &     --          &     1.488       &     70.766      &     0.22423     &     0.53        &     22.6        &              -     \\   
05190\_4     &             -     &         254.295195930     &         -52.770167681     &     21.301      &     --          &     --          &     --          &     70.715      &     0.23070     &     0.36        &     17.7        &              -     \\
\hline \hline
\end{tabular}
}
\label{tab_app_EW}
\end{flushleft}
\end{table*}
\setcounter{table}{1}
\begin{table*}
\begin{flushleft}
 \caption{\footnotesize  continued.}
\resizebox{0.99\textwidth}{!}{
\begin{tabular}{lr ll cccc clcr r}
\hline
\hline
\noalign{\smallskip}
\multicolumn{1}{c}{Star}& \multicolumn{1}{c}{ID} &
\multicolumn{1}{c}{$\alpha_{2000}$}&
\multicolumn{1}{c}{$\delta_{2000}$}&
\multicolumn{1}{c}{$R_{\rm max}$}&
\multicolumn{1}{c}{\ovv}&
\multicolumn{1}{c}{\ovbv}&
\multicolumn{1}{c}{\ovvi}&
\multicolumn{1}{c}{$T_{\rm min}$}&
\multicolumn{1}{c}{Period}&
\multicolumn{1}{c}{Ampl.}&
\multicolumn{1}{c}{Distance}&
\multicolumn{1}{l}{MP}\\
& &\multicolumn{1}{c}{[deg]} &\multicolumn{1}{c}{[deg]} &\multicolumn{1}{c}{[mag]} &\multicolumn{1}{c}{[mag]} &\multicolumn{1}{c}{[mag]} &\multicolumn{1}{c}{[mag]}
&\multicolumn{1}{c}{[HJD]} &\multicolumn{1}{c}{[d]} &\multicolumn{1}{c}{[$R$ mag]} &\multicolumn{1}{c}{[arcmin]} &\multicolumn{1}{c}{\%}\\
\hline
\noalign{\smallskip}
15906\_5     &        107493     &         254.316902665     &         -53.047023021     &     19.467      &     20.264      &     --          &     1.244       &     70.605      &     0.28511     &     0.18        &     26.1        &              -     \\
09914\_1     &          7211     &         254.964750204     &         -52.662654779     &     17.972      &     18.661      &     1.247       &     1.147       &     70.582      &     0.28650     &     0.17        &     7.6         &              -     \\
00564\_2     &         23439     &         254.689745025     &         -52.812725870     &     20.204      &     20.735      &     1.101       &     1.145       &     70.586      &     0.28755     &     0.07        &     6.9         &              -     \\
00920\_4     &         71955     &         254.346375967     &         -52.744068157     &     20.395      &     21.021      &     1.219       &     1.253       &     70.859      &     0.28946     &     0.20        &     15.6        &              -     \\
\hline \hline
\end{tabular}
}
\end{flushleft}
\end{table*}
\setcounter{table}{1}
\begin{table*}
\begin{flushleft}
 \caption{\footnotesize  continued.}
\resizebox{0.99\textwidth}{!}{
\begin{tabular}{lr ll cccc clcr r}
\hline
\hline
\noalign{\smallskip}
\multicolumn{1}{c}{Star}& \multicolumn{1}{c}{ID} &
\multicolumn{1}{c}{$\alpha_{2000}$}&
\multicolumn{1}{c}{$\delta_{2000}$}&
\multicolumn{1}{c}{$R_{\rm max}$}&
\multicolumn{1}{c}{\ovv}&
\multicolumn{1}{c}{\ovbv}&
\multicolumn{1}{c}{\ovvi}&
\multicolumn{1}{c}{$T_{\rm min}$}&
\multicolumn{1}{c}{Period}&
\multicolumn{1}{c}{Ampl.}&
\multicolumn{1}{c}{Distance}&
\multicolumn{1}{l}{MP}\\
& &\multicolumn{1}{c}{[deg]} &\multicolumn{1}{c}{[deg]} &\multicolumn{1}{c}{[mag]} &\multicolumn{1}{c}{[mag]} &\multicolumn{1}{c}{[mag]} &\multicolumn{1}{c}{[mag]}
&\multicolumn{1}{c}{[HJD]} &\multicolumn{1}{c}{[d]} &\multicolumn{1}{c}{[$R$ mag]} &\multicolumn{1}{c}{[arcmin]} &\multicolumn{1}{c}{\%}\\
\hline
\noalign{\smallskip}
12347\_4     &         80409     &         254.338135370     &         -52.562761802     &     18.107      &     19.099      &     0.616       &     1.694       &     70.688      &     0.33057     &     0.46        &     18.0        &              -     \\
11806\_5     &        104429     &         254.343834208     &         -53.030153628     &     20.690      &     21.442      &     1.324       &     1.082       &     70.746      &     0.33164     &     0.25        &     24.7        &              -     \\
15757\_3     &         58488     &         254.492893426     &         -52.687986583     &     18.939      &     19.722      &     0.996       &     1.607       &     70.871      &     0.33170     &     0.22        &     10.2        &              -     \\
14706\_7     &        156370     &         254.680160208     &         -52.879691196     &     18.674      &     19.302      &     0.889       &     1.146       &     70.836      &     0.33554     &     0.08        &     10.8        &              -     \\
\hline \hline
\end{tabular}
}
\end{flushleft}
\end{table*}
\setcounter{table}{1}
\begin{table*}
\begin{flushleft}
 \caption{\footnotesize  continued.}
\resizebox{0.99\textwidth}{!}{
\begin{tabular}{lr ll cccc clcr r}
\hline
\hline
\noalign{\smallskip}
\multicolumn{1}{c}{Star}& \multicolumn{1}{c}{ID} &
\multicolumn{1}{c}{$\alpha_{2000}$}&
\multicolumn{1}{c}{$\delta_{2000}$}&
\multicolumn{1}{c}{$R_{\rm max}$}&
\multicolumn{1}{c}{\ovv}&
\multicolumn{1}{c}{\ovbv}&
\multicolumn{1}{c}{\ovvi}&
\multicolumn{1}{c}{$T_{\rm min}$}&
\multicolumn{1}{c}{Period}&
\multicolumn{1}{c}{Ampl.}&
\multicolumn{1}{c}{Distance}&
\multicolumn{1}{l}{MP}\\
& &\multicolumn{1}{c}{[deg]} &\multicolumn{1}{c}{[deg]} &\multicolumn{1}{c}{[mag]} &\multicolumn{1}{c}{[mag]} &\multicolumn{1}{c}{[mag]} &\multicolumn{1}{c}{[mag]}
&\multicolumn{1}{c}{[HJD]} &\multicolumn{1}{c}{[d]} &\multicolumn{1}{c}{[$R$ mag]} &\multicolumn{1}{c}{[arcmin]} &\multicolumn{1}{c}{\%}\\
\hline
\noalign{\smallskip}
22442\_6     &        136638     &         254.412274265     &         -52.852975207     &     16.488      &     17.114      &     0.835       &     1.064       &     70.859      &     0.39560     &     0.17        &     15.6        &              -     \\
26667\_1     &         18889     &         254.885399844     &         -52.742269102     &     18.648      &     19.550      &     --          &     1.361       &     70.867      &     0.39570     &     0.58        &     4.6         &              -     \\
09705\_5     &        102913     &         254.311160614     &         -52.863616385     &     17.733      &     18.411      &     0.989       &     1.028       &     70.766      &     0.39580     &     0.20        &     19.1        &              -     \\
19141\_1     &         13846     &         254.885985663     &         -52.742281281     &     18.067      &     18.581      &     0.668       &     1.034       &     70.867      &     0.39600     &     0.12        &     4.7         &              -     \\
\hline \hline
\end{tabular}
}
\end{flushleft}
\end{table*}
\begin{table*}
\begin{flushleft}
 \caption{\footnotesize Detached systems, $\beta$ Per (Algol)  type (EA) systems.}
\resizebox{0.99\textwidth}{!}{
\begin{tabular}{lr ll cccc clc r r }
\hline \hline
\noalign{\smallskip}
\multicolumn{1}{c}{Star}& \multicolumn{1}{c}{ID} &
\multicolumn{1}{c}{$\alpha_{2000}$}&
\multicolumn{1}{c}{$\delta_{2000}$}&
\multicolumn{1}{c}{$R_{\rm max}$}&
\multicolumn{1}{c}{\ovv}&
\multicolumn{1}{c}{\ovbv}&
\multicolumn{1}{c}{\ovvi}&
\multicolumn{1}{c}{$T_{\rm min}$}&
\multicolumn{1}{c}{Period}&
\multicolumn{1}{c}{Ampl.}&
\multicolumn{1}{c}{Distance}&
\multicolumn{1}{l}{MP} \\
& &\multicolumn{1}{c}{[deg]} &\multicolumn{1}{c}{[deg]} &\multicolumn{1}{c}{[mag]} &\multicolumn{1}{c}{[mag]} &\multicolumn{1}{c}{[mag]} &\multicolumn{1}{c}{[mag]}
&\multicolumn{1}{c}{[HJD]} &\multicolumn{1}{c}{[d]} &\multicolumn{1}{c}{[$R$ mag]} &\multicolumn{1}{c}{[arcmin]} &\multicolumn{1}{c}{\%}\\
\hline
\noalign{\smallskip}
21471\_1     &         15395     &         255.017261399     &         -52.673015599     &     21.199      &     --          &     --          &     --          &     70.648      &     0.25051     &     0.86        &     9.2         &              -     \\   
21839\_2     &             -     &         254.800351011     &         -52.691146980     &     22.995      &     --          &     --          &     --          &     70.598      &     0.25290     &     1.33        &     1.5         &              -     \\   
25622\_1     &         18191     &         255.016637408     &         -52.559069751     &     20.972      &     21.863      &     1.266       &     1.657       &     70.621      &     0.26185     &     0.51        &     12.7        &              -     \\   
06097\_8     &        169599     &         254.909122586     &         -52.998979068     &     20.318      &     21.536      &     --          &     1.888       &     70.844      &     0.27033     &     0.77        &     18.1        &              -     \\
\hline \hline
\end{tabular}
}
\label{tab_app_EA}
\end{flushleft}
\end{table*}

\begin{table*}
\begin{flushleft}
 \caption{\footnotesize Semi--detached systems, $\beta$ Lyr type (EB) systems.}
\resizebox{0.99\textwidth}{!}{
\begin{tabular}{lr ll cccc clcr r }
\hline \hline
\noalign{\smallskip}
\multicolumn{1}{c}{Star}& \multicolumn{1}{c}{ID} &
\multicolumn{1}{c}{$\alpha_{2000}$}&
\multicolumn{1}{c}{$\delta_{2000}$}&
\multicolumn{1}{c}{$R_{\rm max}$}&
\multicolumn{1}{c}{\ovv}&
\multicolumn{1}{c}{\ovbv}&
\multicolumn{1}{c}{\ovvi}&
\multicolumn{1}{c}{$T_{\rm min}$}&
\multicolumn{1}{c}{Period}&
\multicolumn{1}{c}{Ampl.}&
\multicolumn{1}{c}{Distance}&
\multicolumn{1}{l}{MP} \\
& &\multicolumn{1}{c}{[deg]} &\multicolumn{1}{c}{[deg]} &\multicolumn{1}{c}{[mag]} &\multicolumn{1}{c}{[mag]} &\multicolumn{1}{c}{[mag]} &\multicolumn{1}{c}{[mag]}
&\multicolumn{1}{c}{[HJD]} &\multicolumn{1}{c}{[d]} &\multicolumn{1}{c}{[$R$ mag]} &\multicolumn{1}{c}{[arcmin]} &\multicolumn{1}{c}{\%}\\
\noalign{\smallskip}
\hline
\noalign{\smallskip}
00145\_2     &         23102     &         254.732762682     &         -52.818849796     &     16.494      &     17.223      &     1.228       &     1.198       &     70.676      &     0.29110     &     0.26        &     6.8         &         0     \\   
25252\_3     &         66936     &         254.610787192     &         -52.641122153     &     20.364      &     21.152      &     1.477       &     1.395       &     70.867      &     0.30780     &     0.17        &     7.1         &              -     \\   
08123\_2     &         29437     &         254.789360265     &         -52.705813364     &     19.884      &     20.992      &     1.385       &     2.252       &     70.883      &     0.31903     &     0.15        &     0.7         &              -     \\   
26817\_4     &         91189     &         254.184162106     &         -52.661447237     &     17.866      &     18.505      &     1.117       &     1.061       &     70.586      &     0.35828     &     0.25        &     21.5        &              -     \\
\hline \hline
\end{tabular}
}
\label{tab_app_EB}
\end{flushleft}
\end{table*}

\begin{table*}
\begin{flushleft}
 \caption{\footnotesize RS CVn variables, binary systems showing stellar activity in one or both
components.}
\resizebox{0.99\textwidth}{!}{
\begin{tabular}{lr ll cccc clc r r }
\hline \hline
\noalign{\smallskip}
\multicolumn{1}{c}{Star}& \multicolumn{1}{c}{ID} &
\multicolumn{1}{c}{$\alpha_{2000}$}&
\multicolumn{1}{c}{$\delta_{2000}$}&
\multicolumn{1}{c}{$R_{\rm max}$}&
\multicolumn{1}{c}{\ovv}&
\multicolumn{1}{c}{\ovbv}&
\multicolumn{1}{c}{\ovvi}&
\multicolumn{1}{c}{$T_{\rm min}$}&
\multicolumn{1}{c}{Period}&
\multicolumn{1}{c}{Ampl.}&
\multicolumn{1}{c}{Distance}&
\multicolumn{1}{l}{MP} \\
& &\multicolumn{1}{c}{[deg]} &\multicolumn{1}{c}{[deg]} &\multicolumn{1}{c}{[mag]} &\multicolumn{1}{c}{[mag]} &\multicolumn{1}{c}{[mag]} &\multicolumn{1}{c}{[mag]}
&\multicolumn{1}{c}{[HJD]} &\multicolumn{1}{c}{[d]} &\multicolumn{1}{c}{[$R$ mag]} &\multicolumn{1}{c}{[arcmin]} &\multicolumn{1}{c}{\%}\\
\hline
\noalign{\smallskip}
25516\_5     &        114763     &         254.273572732     &         -53.026533926     &     16.260      &     16.866      &     0.838       &     1.090       &     70.855      &     --          &     0.03        &     26.2        &              -     \\   
02582\_7     &        145527     &         254.864316020     &         -53.055976199     &     18.615      &     19.343      &     1.072       &     1.358       &     --          &     --          &     0.08        &     21.1        &              -     \\   
20346\_3     &         62538     &         254.569287024     &         -52.559414721     &     19.054      &     19.821      &     1.257       &     1.399       &     70.844      &     0.47060     &     0.06        &     11.6        &              -     \\   
06395\_3     &         50575     &         254.482527337     &         -52.591344638     &     17.001      &     17.755      &     1.197       &     1.326       &     70.871      &     0.55286     &     0.06        &     12.6        &              -     \\
\hline \hline
\end{tabular}
}
\label{tab_app_RSCVn}
\end{flushleft}
\end{table*}

\begin{table*}
\begin{flushleft}
 \caption{\footnotesize Single--wave rotational variables (RO1).}
\resizebox{0.99\textwidth}{!}{
\begin{tabular}{lr ll cccc clcc r }
\hline \hline
\noalign{\smallskip}
\multicolumn{1}{c}{Star}& \multicolumn{1}{c}{ID} &
\multicolumn{1}{c}{$\alpha_{2000}$}&
\multicolumn{1}{c}{$\delta_{2000}$}&
\multicolumn{1}{c}{\ovr}&
\multicolumn{1}{c}{\ovv}&
\multicolumn{1}{c}{\ovbv}&
\multicolumn{1}{c}{\ovvi}&
\multicolumn{1}{c}{$T_{\rm min}$}&
\multicolumn{1}{c}{Period}&
\multicolumn{1}{c}{Ampl.}&
\multicolumn{1}{c}{Distance}&
\multicolumn{1}{l}{MP} \\
& &\multicolumn{1}{c}{[deg]} &\multicolumn{1}{c}{[deg]} &\multicolumn{1}{c}{[mag]} &\multicolumn{1}{c}{[mag]} &\multicolumn{1}{c}{[mag]} &\multicolumn{1}{c}{[mag]}
&\multicolumn{1}{c}{[HJD]} &\multicolumn{1}{c}{[d]} &\multicolumn{1}{c}{[$R$ mag]} &\multicolumn{1}{c}{[arcmin]} &\multicolumn{1}{c}{\%}\\
\hline
\noalign{\smallskip}
02849\_5     &         98253     &         254.354170197     &         -52.871187225     &     14.559      &     15.082      &     0.792       &     0.989       &     70.805      &     0.34014     &     0.03        &     18.0        &              -     \\   
08107\_3     &         51940     &         254.473655174     &         -52.562349360     &     18.548      &     19.122      &     0.908       &     1.132       &     70.648      &     0.34119     &     0.01        &     13.9        &              -     \\   
26178\_5     &        115311     &         254.352916128     &         -52.832609795     &     19.503      &     19.605      &     0.304       &     0.474       &     70.723      &     0.37428     &     0.08        &     16.9        &              -     \\   
27582\_2     &             -     &         254.733590738     &         -52.598549138     &     21.301      &     21.738      &     1.064       &     1.359       &     70.762      &     0.35688     &     0.34        &     6.7         &              -     \\
\hline \hline
\end{tabular}
}
\label{tab_app_RO1}
\end{flushleft}
\end{table*}

\begin{table*}
\begin{flushleft}
 \caption{\footnotesize Double--wave rotational variables (RO2).}
\resizebox{0.99\textwidth}{!}{
\begin{tabular}{lr ll cccc clcr r }
\hline \hline
\noalign{\smallskip}
\multicolumn{1}{c}{Star}& \multicolumn{1}{c}{ID} &
\multicolumn{1}{c}{$\alpha_{2000}$}&
\multicolumn{1}{c}{$\delta_{2000}$}&
\multicolumn{1}{c}{\ovr}&
\multicolumn{1}{c}{\ovv}&
\multicolumn{1}{c}{\ovbv}&
\multicolumn{1}{c}{\ovvi}&
\multicolumn{1}{c}{$T_{\rm min}$}&
\multicolumn{1}{c}{Period}&
\multicolumn{1}{c}{Ampl.}&
\multicolumn{1}{c}{Distance}&
\multicolumn{1}{l}{MP} \\
& & \multicolumn{1}{c}{[deg]} &\multicolumn{1}{c}{[deg]} &\multicolumn{1}{c}{[mag]} &\multicolumn{1}{c}{[mag]} &\multicolumn{1}{c}{[mag]} &\multicolumn{1}{c}{[mag]}
&\multicolumn{1}{c}{[HJD]} &\multicolumn{1}{c}{[d]} &\multicolumn{1}{c}{[$R$ mag]} &\multicolumn{1}{c}{[arcmin]} &\multicolumn{1}{c}{\%}\\
\hline
\noalign{\smallskip}
12912\_3     &         56017     &         254.533555002     &         -52.784475843     &     19.658      &     20.444      &     0.990       &     1.347       &     70.750      &     0.20639     &     0.07        &     9.8         &              -     \\   
04317\_3     &         49008     &         254.608573445     &         -52.794792573     &     17.503      &     18.035      &     0.961       &     1.053       &     70.891      &     0.32733     &     0.03        &     7.8         &              -     \\   
07671\_4     &         76978     &         254.335531957     &         -52.705889637     &     17.150      &     17.571      &     0.729       &     0.901       &     70.668      &     0.61286     &     0.03        &     15.8        &              -     \\   
06474\_2     &         28156     &         254.756128298     &         -52.729271142     &     18.575      &     19.186      &     1.051       &     1.363       &     70.949      &     0.89900     &     0.06        &     1.4         &         9*     \\
\hline \hline
\end{tabular}
}
\label{tab_app_RO2}
\end{flushleft}
\end{table*}
\clearpage

\longtab{8}{
\begin{longtable}{lr ll cccc cr r}
 \caption{\label{tab_app_LON} Long-period variables.}\\
\hline
\hline
\noalign{\smallskip}
\multicolumn{1}{c}{Star}& \multicolumn{1}{c}{ID} &
\multicolumn{1}{c}{$\alpha_{2000}$}&
\multicolumn{1}{c}{$\delta_{2000}$}&
\multicolumn{1}{c}{\ovr}&
\multicolumn{1}{c}{\ovv}&
\multicolumn{1}{c}{\ovbv}&
\multicolumn{1}{c}{\ovvi}&
\multicolumn{1}{c}{Ampl.}&
\multicolumn{1}{c}{Distance}&
\multicolumn{1}{l}{MP}\\
& &\multicolumn{1}{c}{[deg]} &\multicolumn{1}{c}{[deg]} &\multicolumn{1}{c}{[mag]} &\multicolumn{1}{c}{[mag]} &\multicolumn{1}{c}{[mag]} &\multicolumn{1}{c}{[mag]}
&\multicolumn{1}{c}{[$R$ mag]} &\multicolumn{1}{c}{[arcmin]} &\multicolumn{1}{c}{\%}\\
\noalign{\smallskip}
\hline
\noalign{\smallskip}
\endfirsthead
\caption{continued.}\\
\hline\hline
\noalign{\smallskip}
\multicolumn{1}{c}{Star}& \multicolumn{1}{c}{ID} &
\multicolumn{1}{c}{$\alpha_{2000}$}&
\multicolumn{1}{c}{$\delta_{2000}$}&
\multicolumn{1}{c}{\ovr}&
\multicolumn{1}{c}{\ovv}&
\multicolumn{1}{c}{\ovbv}&
\multicolumn{1}{c}{\ovvi}&
\multicolumn{1}{c}{Ampl.}&
\multicolumn{1}{c}{Distance}&
\multicolumn{1}{l}{MP} \\
&  &\multicolumn{1}{c}{[deg]} &\multicolumn{1}{c}{[deg]} &\multicolumn{1}{c}{[mag]} &\multicolumn{1}{c}{[mag]} &\multicolumn{1}{c}{[mag]} &\multicolumn{1}{c}{[mag]}
&\multicolumn{1}{c}{[$R$ mag]} &\multicolumn{1}{c}{[arcmin]} &\multicolumn{1}{c}{\%}\\
\noalign{\smallskip}
\hline
\noalign{\smallskip}
\endhead
\hline
\endfoot
15877\_2     &             -     &         254.703470476     &         -52.602380112     &     23.035      &       -         &       -         &       -
   &     3.0:        &     6.8         &              -     \\
11351\_8     &        173273     &         254.896720660     &         -52.836752470     &     14.465      &     15.205      &     1.285       &     1.523       &     0.07        &     9.0         &              -     \\   
26313\_3     &         67881     &         254.415548223     &         -52.753401371     &     13.999      &     15.261      &     1.798       &     2.677       &     0.04        &     13.2        &              -     \\   
05327\_6     &        124207     &         254.603157998     &         -53.029501689     &     14.850      &     15.324      &     0.881       &     0.957       &     0.03        &     20.2        &              -     \\   
04281\_3     &         48982     &         254.572575855     &         -52.797861044     &     14.531      &     15.331      &     1.279       &     1.495       &     0.02        &     9.0         &              -     \\   
04319\_3     &         49009     &         254.427392028     &         -52.794585896     &     14.588      &     15.372      &     1.254       &     1.491       &     0.03        &     13.5        &              -     \\   
23474\_5     &        113269     &         254.198230796     &         -53.082982900     &     14.599      &     15.400      &     1.471       &     1.531       &     0.09        &     30.6        &              -     \\   
03878\_4     &         74227     &         254.188239396     &         -52.805213798     &     14.630      &     15.417      &     1.169       &     1.485       &     0.05        &     21.9        &              -     \\
\hline
\hline
\end{longtable}
}

\begin{table*}[]
\begin{flushleft}
\caption{ \footnotesize  List of binary systems (contact, detached and semi-detached  binaries)
located at less than 8\arcmin~ from the centre
which are considered as candidate members of NGC~6253.}
\resizebox{0.99\textwidth}{!}{
\begin{tabular}{lrl ll cccc clcc r cccc}
\hline \hline
\noalign{\smallskip}
& & & & & & & & & & & & & & \multicolumn{4}{c}{Membership}\\
\multicolumn{1}{c}{Star}& \multicolumn{1}{c}{ID} &
\multicolumn{1}{c}{Type}&
\multicolumn{1}{c}{$\alpha_{2000}$}&
\multicolumn{1}{c}{$\delta_{2000}$}&
\multicolumn{1}{c}{$R_{\rm max}$}&
\multicolumn{1}{c}{\ovv}&
\multicolumn{1}{c}{\ovbv}&
\multicolumn{1}{c}{\ovvi}&
\multicolumn{1}{c}{$T_{\rm min}$}&
\multicolumn{1}{c}{Period}&
\multicolumn{1}{c}{Ampl.}&
\multicolumn{1}{c}{Distance}&
\multicolumn{1}{l}{MP}
& \multicolumn{1}{c}{CMD} & \multicolumn{1}{c}{CMD} & \multicolumn{1}{c}{PLC} & \multicolumn{1}{c}{PLC} \\
& & &\multicolumn{1}{c}{[deg]} &\multicolumn{1}{c}{[deg]} &\multicolumn{1}{c}{[mag]} &\multicolumn{1}{c}{[mag]} &\multicolumn{1}{c}{[mag]} &\multicolumn{1}{c}{[mag]}
&\multicolumn{1}{c}{[HJD]} &\multicolumn{1}{c}{[d]} &\multicolumn{1}{c}{[$R$ mag]} &\multicolumn{1}{c}{[arcmin]} &\multicolumn{1}{c}{\%}
&  \multicolumn{1}{c}{B-V} & \multicolumn{1}{c}{V-I} & \multicolumn{1}{c}{B-V} & \multicolumn{1}{c}{V-I} \\
\hline
\noalign{\smallskip}
\multicolumn{14}{c}{Contact binaries}\\
\noalign{\smallskip}
23188\_2 & 41404  &EW&  254.656188433 &   -52.61619778  &    14.464   &  14.928   &  0.751     &  0.923     &  70.594    &  0.49680   &  0.32      &  6.9       &  87  & y & y & y & y   \\
10853\_2 & 31606  &EW&  254.800782797 &   -52.669356128 &    16.342   &  16.921   &  1.049     &  1.191     &  70.777    &  0.29340   &  0.02      &  2.6       &  78  & y & y & y & y   \\
01015\_2 & 23797  &EW&  254.782447929 &   -52.805883546 &    15.462   &  15.969   &  0.874     &  1.036     &  70.637    &  0.38720   &  0.05      &  5.9       &  22  & y & y & y & y   \\
13288\_2 & 33521  &EW&  254.673007553 &   -52.637555967 &    14.807   &  15.444   &  1.291     &  0.955     &  70.867    &  0.47290   &  0.44      &  5.5       &  3   & n & y & y & y   \\
\hline \hline
\end{tabular}
}
\label{tab_EW_n6253}
\end{flushleft}
\end{table*}

\begin{table*}[]
\caption{\footnotesize   List of rotational and long-period variables located at less
 than 8\arcmin\, from the centre which are considered to be candidate cluster members.}
\begin{flushleft}
\resizebox{0.99\textwidth}{!}{
\begin{tabular}{lrl ll cccc clcc r }
\hline \hline
\noalign{\smallskip}
\multicolumn{1}{c}{Star}& \multicolumn{1}{c}{ID} &
\multicolumn{1}{c}{Type}&
\multicolumn{1}{c}{$\alpha_{2000}$}&
\multicolumn{1}{c}{$\delta_{2000}$}&
\multicolumn{1}{c}{\ovr}&
\multicolumn{1}{c}{\ovv}&
\multicolumn{1}{c}{\ovbv}&
\multicolumn{1}{c}{\ovvi}&
\multicolumn{1}{c}{$T_{\rm max}$}&
\multicolumn{1}{c}{Period}&
\multicolumn{1}{c}{Ampl.}&
\multicolumn{1}{c}{Distance}&
\multicolumn{1}{l}{MP} \\
&&  &\multicolumn{1}{c}{[deg]} &\multicolumn{1}{c}{[deg]} &\multicolumn{1}{c}{[mag]} &\multicolumn{1}{c}{[mag]} &\multicolumn{1}{c}{[mag]} &\multicolumn{1}{c}{[mag]}
&\multicolumn{1}{c}{[HJD]} &\multicolumn{1}{c}{[d]} &\multicolumn{1}{c}{[$R$ mag]} &\multicolumn{1}{c}{[arcmin]} &\multicolumn{1}{c}{\%}\\
\hline
\noalign{\smallskip}
09901\_2   &    30842 & RO1 &   254.732275735 &   -52.681556393  &    17.619    & 18.247    &  1.038  & 1.351 &      72.668     &  1.11842   &   0.08   &  2.1  & ---  \\
11977\_1   &     8713 & RO1 &   254.929865891 &   -52.630746475  &    16.185    & 16.592    &  0.749  & 0.849 &      71.188     &  1.8424    &   0.01   &  7.4  & ---  \\
07392\_1   &     5378 & RO1 &   254.904680899 &   -52.702647524  &    17.641    & 18.217    &  1.040  & 1.195 &      70.914     &  2.982     &   0.11   &  4.9  & ---  \\
14911\_2   &    34814 & RO1 &   254.721536966 &   -52.615435459  &    17.702    & 18.299    &  1.053  & 1.212 &      70.902     &  1.5806    &   0.05   &  5.9  &  63  \\
10372\_2   &    31220 & RO1 &   254.679930555 &   -52.675735040  &    17.650    & 18.317    &  1.080  & 1.456 &      72.957     &  2.735     &   0.04   &  3.8  &  26  \\
15993\_2   &    35676 & RO1 &   254.846280438 &   -52.600497324  &    18.851    & 19.741    &  1.462  & 1.844 &      71.766     &  3.151     &   0.03   &  7.0  & ---  \\
\hline \hline
\end{tabular}
}
\end{flushleft}
\label{tab_rot_n6253}
\end{table*}

\end{appendix}
\clearpage

\begin{figure*}
\includegraphics[scale=0.9]{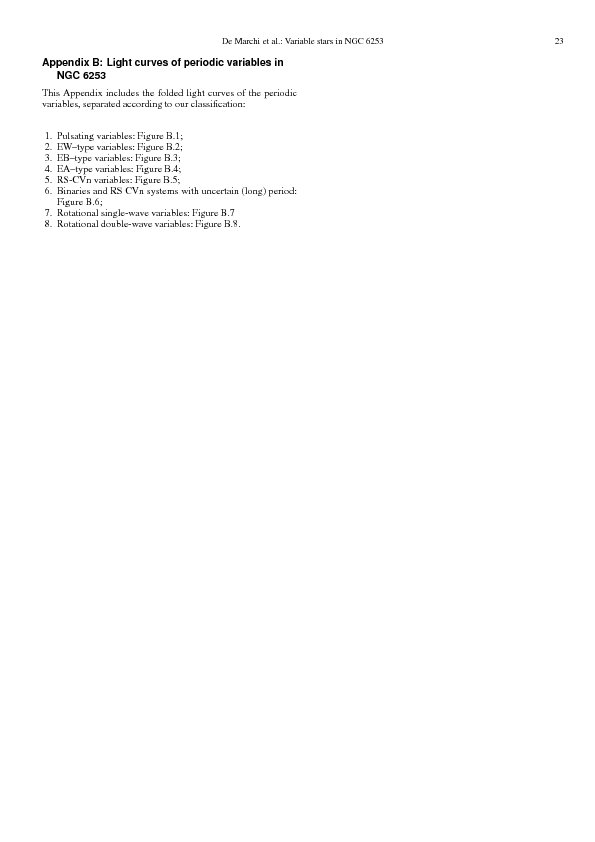}
\end{figure*}

\begin{figure*}
\includegraphics[scale=0.9]{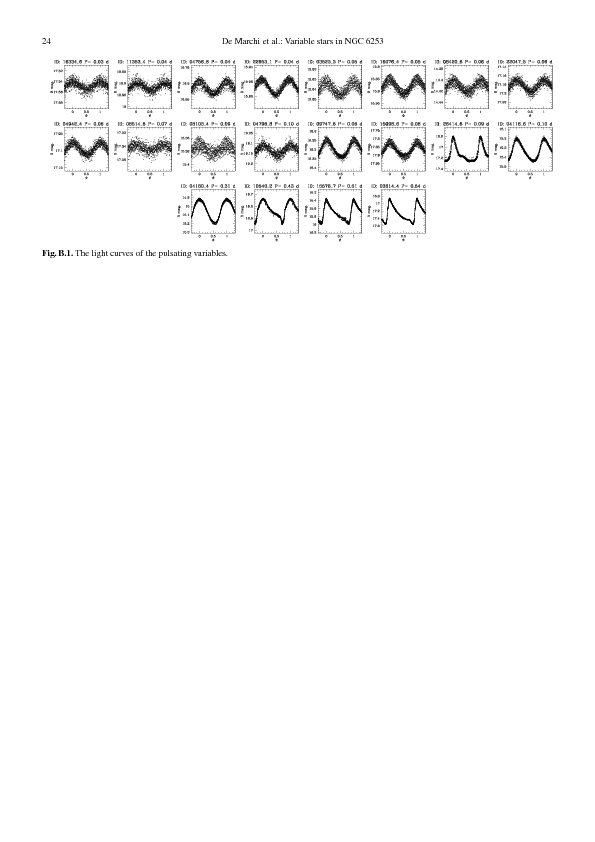}
\end{figure*}

\begin{figure*}
\includegraphics[scale=0.9]{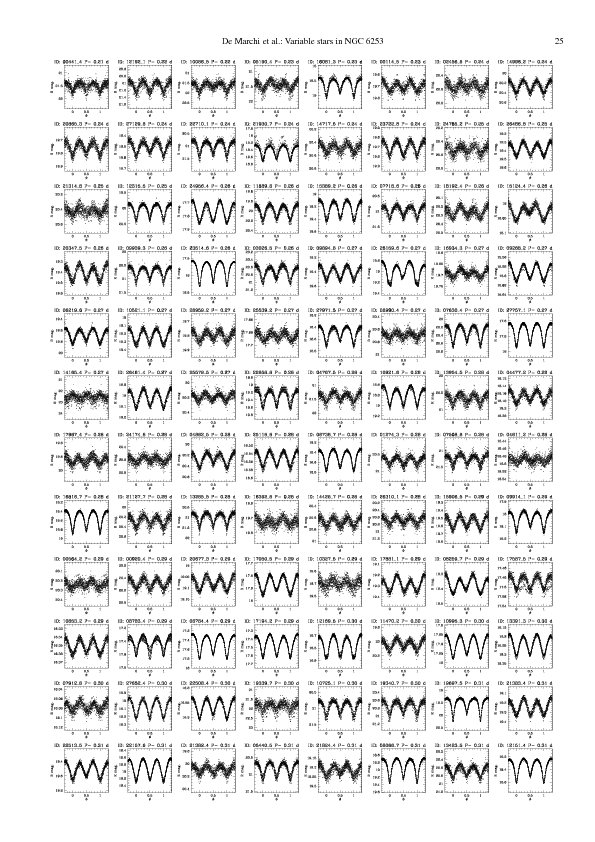}
\end{figure*}
\begin{figure*}
\includegraphics[scale=0.9]{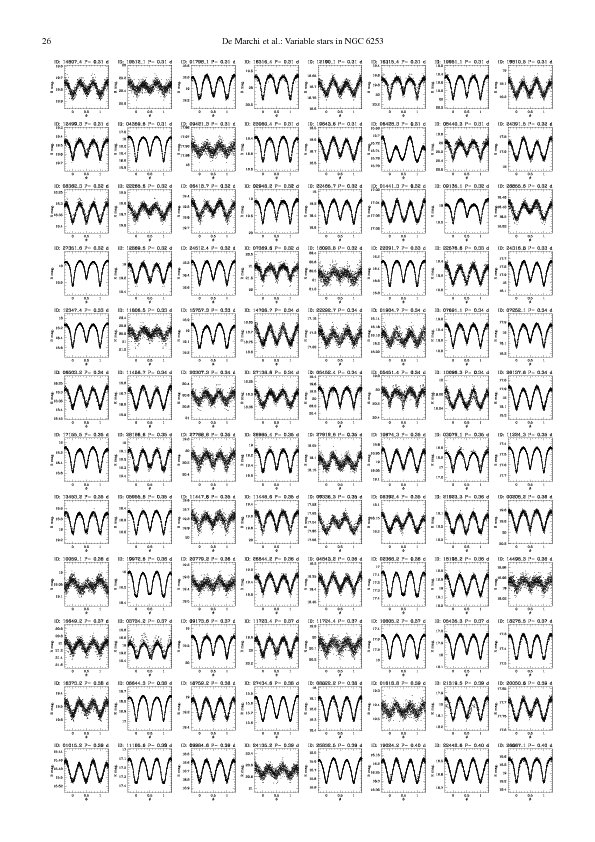}
\end{figure*}
\begin{figure*}
\includegraphics[scale=0.9]{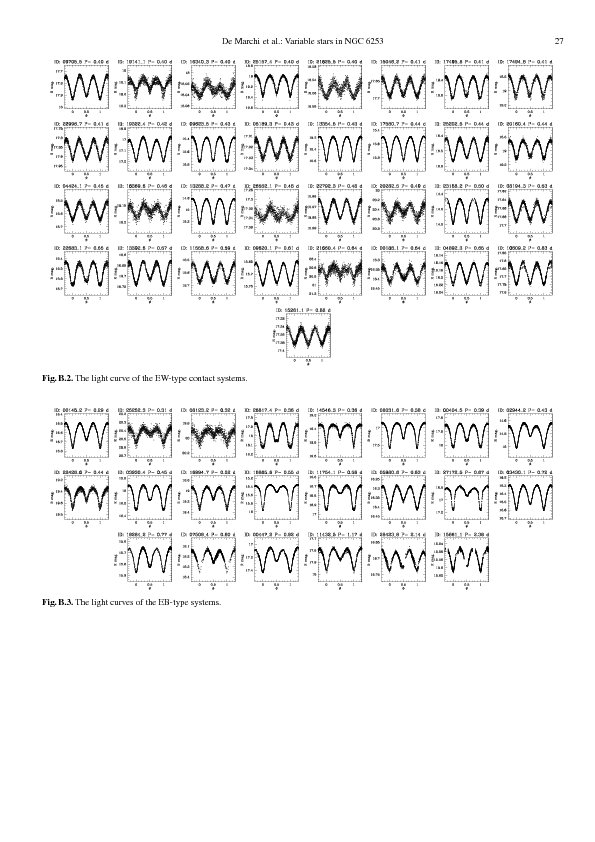}
\end{figure*}
\begin{figure*}
\includegraphics[scale=0.9]{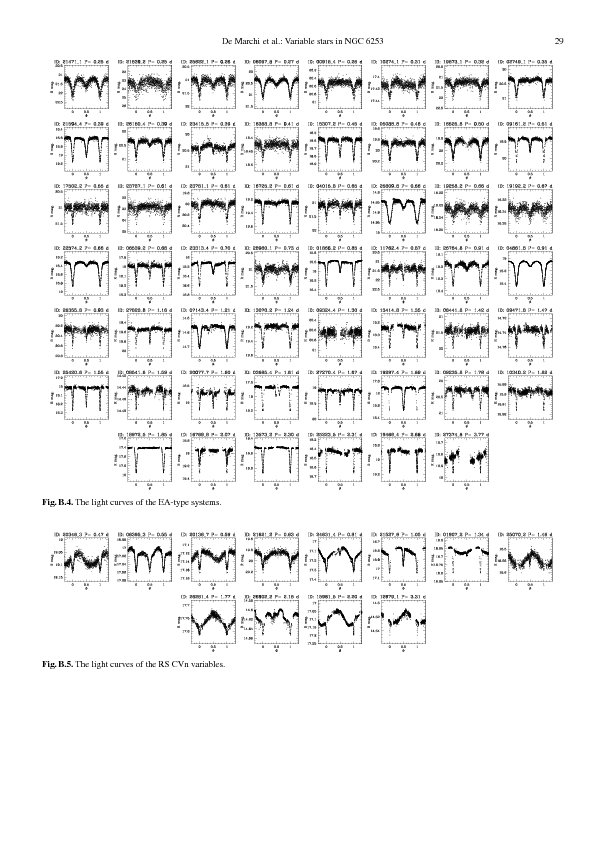}
\end{figure*}
\begin{figure*}
\includegraphics[scale=0.9]{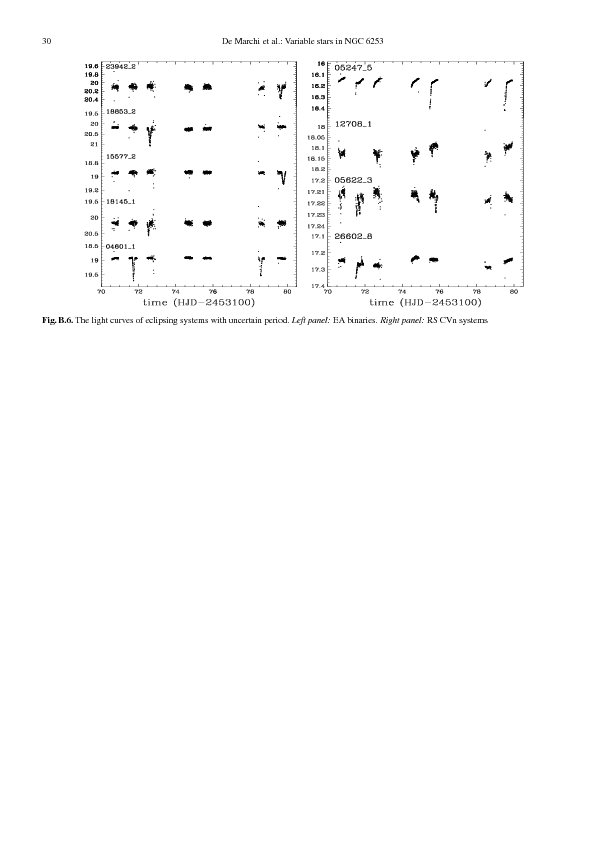}
\end{figure*}
\begin{figure*}
\includegraphics[scale=0.9]{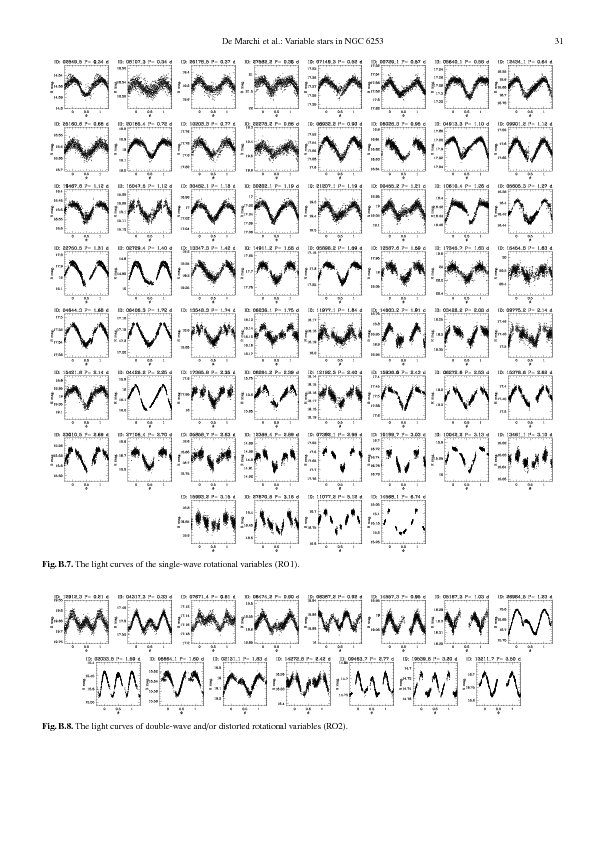}
\end{figure*}

\end{document}